\documentclass{Interspeech2024}
\usepackage{multirow}

\interspeechcameraready

\title{Exploring Spoken Language Identification Strategies for Automatic Transcription of Multilingual Broadcast and Institutional Speech}

\name[affiliation={1}]{Martina}{Valente}
\name[affiliation={1}]{Fabio}{Brugnara}
\name[affiliation={1}]{Giovanni}{Morrone}
\name[affiliation={1}]{Enrico}{Zovato}
\name[affiliation={1}]{Leonardo}{Badino}

\address{
  $^1$Almawave S.p.A., Voice Engineering Lab, Italy}
\email{\{m.valente,f.brugnara,g.morrone,e.zovato,l.badino\}@almawave.it}


\begin{document}

\maketitle

\begin{abstract}
    This paper addresses spoken language identification (SLI) and speech recognition of multilingual broadcast and institutional speech, real application scenarios that have been rarely addressed in the SLI literature. 
    Observing that in these domains language changes are mostly associated with speaker changes, we propose a cascaded system consisting of speaker diarization and language identification and compare it with more traditional language identification and language diarization systems. 
    Results show that the proposed system often achieves lower language classification and language diarization error rates (up to 10\% relative language diarization error reduction and 60\% relative language confusion reduction) and 
    leads to lower WERs on multilingual test sets (more than 8\% relative WER reduction), while at the same time does not negatively affect speech recognition on monolingual audio (with an absolute WER increase between 0.1\% and 0.7\% w.r.t. monolingual ASR).     
\end{abstract}

\section{Introduction}

Spoken Language Identification (SLI) refers to the task of automatically recognizing the language spoken in a given utterance and is an important preprocessing step for various applications, including Automatic Speech Recognition (ASR).

Most of the SLI systems that have been proposed over the years focus on automatically recognizing the language of short, monolingual audio clips \cite{li2013}, and most publicly available multilingual datasets adhere to this specification \cite{valk2020voxlingua,ardila2020,conneau2023}. Only few studies have approached the challenge of language identification of authentic multilingual audios, most of which target code-switching speech \cite{lyu2013, barras2020, liu21_interspeech}, which is a type of verbal communication in which multiple languages are alternated within a single sentence. In addition, existing works rarely consider the impact that a SLI front-end might have on subsequent processing stages, such as ASR. 

This paper focuses on SLI and its impact on speech recognition of multilingual audio, specifically targeting real application scenarios of broadcast (i.e., speech from TV and radio channels) and institutional speech. In these particular scenarios, which are rather unexplored in the literature, language changes happen at a relatively long timescale and typically match speaker changes. We propose a speaker-informed approach to SLI that combines speaker change detection and SLI and compare it against speaker-agnostic systems that are by design more suitable for more popular scenarios in the SLI literature (e.g. segment-based language classification or code-switching). Other than a simple addition of information, speaker change detection might represent a more accurate proxy for language change detection, given that speaker information requires less context to be extracted with respect to language information. Please note that the proposed technique is not intended to address code-switching by the same speaker.

Stemming from the NeMo Titanet-LID \cite{jia23b_interspeech} trained on VoxLingua107 \cite{valk2020voxlingua}, we train two different SLI architectures: one that assigns language labels to segments identified by previous processing blocks (segment-based) and another one that predicts language labels at fixed time steps (frame-based).

For the first model, we re-use the same architecture as the pretrained model, and fine-tune on domain-specific data to improve the SLI performance on the domain and languages of interest. At inference time, the model classifies the segments identified by either a voice activity detection (VAD) or a speaker diarization (SD) system.

For the second model, we extend and finetune the pretrained model architecture to produce frame-based language predictions. We aim at performing both language segmentation and identification from the raw audio, a task known in the code-switching literature as Language Diarization (LD)\cite{lyu2013}. Recent works have shown that end-to-end LD systems that combine local language information extraction with a contextual language classifier are able to outperform more traditional systems based on two-step processing of embedding extraction and classification \cite{liu21_interspeech}. Additionally, the use of pretrained models as feature extractors do increase the identification performance \cite{mishra23_interspeech}. Inspired by this type of considerations, we develop an end-to-end system for language diarization that combines a pretrained convolutional front-end for the extraction of local information with a back-end based on the Long Short-Term Memory (LSTM) topology \cite{graves2005framewise} to take into account the contextual information.

We report extensive evaluations of the different language identification systems on test sets from broadcast and institutional domains and compare our results with a baseline system composed by publicly available VAD and SLI provided by SpeechBrain \cite{speechbrain}.
In addition, we assess the impact of language identification front-end on the transcription accuracy of a multilingual system composed by SLI followed by monolingual speech recognition engines.
We further address the question of whether a SLI is worthwhile even when the input signal is mostly monolingual: this is a relevant question especially in broadcast speech where multilingual speech is present but most of the audio is monolingual. 

\section{System Architecture}

We test different combinations of segmentation and language identification models.
More specifically, we compare the performance of three different cascaded systems: 
\begin{itemize}
    \item Voice Activity Detection followed by segment-based SLI.
    \item Speaker Diarization followed by segment-based SLI.
    \item Voice Activity Detection followed by frame-based SLI.
\end{itemize}

In addition, we combine the language identification systems with monolingual ASR engines to build multilingual speech recognition systems, and test the impact of such a configuration on transcription accuracy.

In the following, we provide the implementation details of each of the components mentioned above.

\subsection{Segment-based and frame-based SLI models}

The segment-based SLI model is a fine-tuned version of the TitaNet-LID\footnote{\url{https://catalog.ngc.nvidia.com/orgs/nvidia/teams/nemo/models/langid_ambernet}} \cite{jia23b_interspeech} trained on VoxLingua107 \cite{valk2020voxlingua}. The TitaNet-LID architecture consists of a 1D depth-wise channel separable convolutional encoder with ContextNet-like architecture \cite{han20_interspeech} and a decoder with a statistic pooling
layer followed by two linear layers. This model shows state-of-the-art results on the VoxLingua107 test set while being 10 times smaller (29M parameters) than competitor models.

For the frame-based SLI model, the TitaNet-LID architecture has been adapted to produce frame-based predictions, by retaining the encoder and replacing the decoder with a bi-LSTM decoder.

\subsection{Speaker Diarization Model}
We employ an improved version of the end-to-end neural diarization (EEND)-vector clustering model \cite{kinoshita21_interspeech}, a hybrid diarization approach that combines neural and clustering-based diarization \cite{park2022review}. The input recording is split into short chunks (e.g., 30 seconds), for which we can assume that a maximum number of active speakers can occur (e.g., 2 or 3). Then, local neural diarization is applied on each chunk independently and outputs speech activity probabilities for each active speaker. For local diarization we use the same multi-layer Transformer-based architecture as in \cite{fujita2019end}. In the original EEND-VC implementation, the same network is exploited to compute speaker embeddings for each active speaker along with diarization results.
Finally, a constrained clustering algorithm is applied on speaker embeddings to estimate which diarization outputs across chunks belong to the same speaker.

We found that the original EEND-VC performed poorly on broadcast audio, suffering the presence of many speakers (e.g., $\ge$ 5) in long input recordings. In such a case, speaker embeddings quality is not good enough to assure decent performance for the following clustering step, thus we employ an external pre-trained embedding extractor \cite{zeinali2019but}.  
This variant of EEND-VC produced a diarization error rate reduction of 49.3\% and 55.9\% on VoxConverse \cite{chung2020spot} dev and test sets, respectively.

In the proposed SLI systems, the SD model is only used to detect speaker changes. Some preliminary experiments using speaker identity to assign language labels led to worse results, but we will explore this aspect more thoroughly in future experiments.

\subsection{Voice Activity Detection Model}
\label{ssec:vad}
To perform a fair comparison, the VAD model is derived from the local neural diarization model. We discard speaker information and only retain speech/non-speech segmentation. The overall VAD results is simply obtained by stitching the local VAD outputs without any clustering.

\section{Experimental Setup}

\subsection{Data}

\subsubsection{Broadcast Datasets}

The broadcast training dataset is composed of 300 hours of audio extracted from TV, radio channels and YouTube. 
It covers 3 languages: Danish, Swedish and English, with approximately 100 hours each. Three different varieties of English were included: British (30 hours), American (30 hours), and L2 English (40 hours). For SLI model training, the audio was split into speech segments of duration between 1 and 20 seconds, and 5\% of the data was used as validation set.

A monolingual test set with approximately 1 hour of material per language was extracted from similar sources as the training and for the same set of languages. The test set was manually reviewed to exclude any foreign language material and transcribed. No annotation of speech/non speech was provided.

A synthetic multilingual test sets was built by extracting monolingual segments of random length from the monolingual test set and concatenating them in random order. Danish, Swedish and English were sampled with uniform probability. We created two versions of the test: \emph{ts-mix-short}, with segment duration from 5 to 15 seconds, and \emph{ts-mix-long} with segments from 15 to 45 seconds, and with two different concatenation modalities: with a gap of one second or without any gap. For each version and concatenation modality, 60 files of around 1 minute were created, resulting in one hour of audio per each version.

The real multilingual test set is a proprietary test set containing extracts from multilingual Swedish TV programs. 
In Scandinavian broadcasting, it is common to find programs that are primarily in one language, e.g. Swedish, featuring interviews or reports in a different language, e.g. English. We collected around 1 hour of material, which was manually annotated with both language labels and verbatim transcripts, for a total of around 40 minutes of Swedish, 10 minutes of English and 1 minute of Danish. The test set counts 76 language changes with monolingual segments length ranging from 0.83 to 362.04 seconds (average: 44.31s).

\subsubsection{Institutional Datasets}
\label{section:institutional_dataset}

The institutional dataset amounts to approximately 1000 hours of material collected from 2008-2020 plenary sessions of the European Parliament\footnote{\url{https://www.europarl.europa.eu/plenary/en/debates-video.html}}. Sessions consist of interventions from multiple speakers in different European languages.  
The transcripts include information about the language of the intervention. We select the 10 most represented languages in the dataset (i.e., English, German, Italian, French, Polish, Spanish, Romanian, Dutch, Greek and Portuguese).
Around 100 hours per language were used for SLI model training, and audio was split into speech segments of duration between 1 and 20 seconds. 3\% of data was used as validation set.

The monolingual test set consists of approximately 10 hours taken from late 2020 plenary sessions of the European Parliament, excluded from the training material. Language coverage matches that of the training set. 
The test set was manually reviewed to exclude non-target language content and to adjust the official transcripts to verbatim standard. 

The multilingual test set 
contains the 10 target languages only. It counts 32 variable-length recordings, containing from 1 to 6 languages each, for an overall 2.6 hours.
Language segmentation and transcription boundaries were manually adjusted. Transcripts were taken from the official source, and are not verbatim. The test set is publicly available \footnote{\url{https://awdelivery.blob.core.windows.net/spoken-language-identification/tst_is24.zip}}. 

\subsubsection{Speaker Diarization Datasets}
Since large broadcast or institutional datasets with diarization labels are not available, we used the \emph{Fisher Corpus Part 1} and \emph{Part 2} \cite{cieri2004fisher} to train the local diarization network.  
After removing problematic annotations, the final training set resulted in 10653 examples, totaling about 1762 hours of English speech sampled at 8 kHz. 
We augmented the dataset following \cite{fujita2019bend} and using
noises from MUSAN corpus \cite{snyder2015musan}, and artificial reverberation \cite{ko2017study}. Through augmentation we generated 4 subsets, each containing 50000 diarization-style simulated mixtures, totaling 8089 hours of speech. In each subset the number of speakers involved is fixed (i.e., from 1 to 4).

For validation and hyperparameters tuning, we used the development set of the VoxConverse dataset \cite{chung2020spot}, which includes broadcast recordings and covers acoustic and linguistic domains very similar to the ones covered in the present work. In particular, the number of speakers involved can range from 1 to 20.

\subsection{Architecture, Training and Inference Details}

\subsubsection{Segment-based and frame-based SLI}

To train SLI models, we used the NeMo toolkit \cite{kuchaiev2019nemo}. We followed the original implementation for feature extraction. For both models, the encoder weights were initialized from the pretrained model. For segment based SLI, the decoder outputs were set to 3 or 10 depending on the experimental domain. Fine-tuning was performed with a learning rate of 5e-5 for 3 epochs and the best checkpoint in terms of validation accuracy was selected. For the frame-based SLI, the decoder was replaced with a 1-layer, 1024 nodes bidirectional LSTM. We tried other configurations, e.g. adding more LSTM layers or removing the squeeze-and-excitation mechanism from the encoder, but these produced slightly worse results. In order to train the model to perform language segmentation, artificial multilingual samples were created on-the-fly by concatenating monolingual samples of length $<$ 10s. The models were trained with a learning rate of 5e-6 for 10 epochs on a Quadro RTX 8000 GPU, and the best checkpoint in terms of validation accuracy was selected.

At inference time, segments shorter than 1s were discarded and segments longer than 20s were split to better match the training conditions. For the frame-based SLI model, we applied a moving average to smooth posterior probabilities, with a window size of 200 frames (i.e., 2 seconds).

\subsubsection{Speaker Diarization}

Our SD system is built on the official implementation\footnote{\url{https://github.com/nttcslab-sp/EEND-vector-clustering}} of the EEND-vector clustering \cite{kinoshita21_interspeech} model. The local diarization neural network consists of self-attentive 6-layer Transformer encoder with 8 attention heads. We basically followed the original paper for feature extraction, network architecture and training protocol with the following exception: the model was trained for 200 epochs on 30 s long chunks of the augmented Fisher training set. The model parameters of the last 10 epochs were averaged to obtain the final model.
We replaced the speaker embedding estimation layer with an external pretrained speaker embedding extractor available in the Wespeaker framework\footnote{\url{https://github.com/wenet-e2e/wespeaker/tree/master/examples/voxceleb/v2}} \cite{wang2023wespeaker}. In particular, we used the 34-layer ResNet based \emph{"r-vector"} architecture \cite{zeinali2019but} trained on VoxCeleb2 \cite{chung2018voxceleb2}. As clustering algorithm we employed the constrained Agglomerative Hierarchical Clustering (AHC) as it reaches the best performance in \cite{kinoshita21_interspeech}.
We tuned the speech activity threshold and the AHC linkage threshold on the VoxConverse dev set.

\subsection{Evaluation metrics and baselines}

For the evaluation of SLI systems, we employ two different performance metrics:
\begin{itemize}
    \item Language Diarization Error Rate (LDER), as defined in \cite{chua23_interspeech}. It is computed as the sum of language confusion (LC), false alarms (FA) and missed speech (MS) normalized by the total audio duration. It takes into account the system-level language identification and segmentation performance. We used this metric for datasets for which we have both language identity and speech/non-speech annotations. 
    \item Language Error Rate (LER), here defined as the ratio of time with incorrect language label to total time recognized as speech. It accounts for language confusion error only. We used this metric for datasets where speech/nonspeech annotation is not available. 
\end{itemize}

We compute 95\% confidence intervals on the LDER measure with bootstrapping approach using the ConfidenceIntervals Python toolkit\footnote{\url{https://github.com/luferrer/ConfidenceIntervals}} applied to frame-level labels.  

We compare our results with two baselines: the first one was obtained by applying SpeechBrain VAD\footnote{\url{https://huggingface.co/speechbrain/vad-crdnn-libriparty}} followed by SpeechBrain ECAPA-TDNN SLI\footnote{\url{https://huggingface.co/speechbrain/lang-id-voxlingua107-ecapahttps://huggingface.co/speechbrain/lang-id-voxlingua107-ecapa}} trained on VoxLingua107 dataset; the second one is the NeMo TitaNet-LID model, that was used to initialize our systems, combined with our in-house VAD (cfr. Section \ref{ssec:vad}). To ensure a fair comparison, we limit both the SpeechBrain and NeMo SLI outputs during inference to the relevant classes by applying a binary mask on the networks' output nodes. For SpeechBrain VAD, we use the default configuration for all parameters.

\section{Results}
\label{section:results}

\subsection{Monolingual Speech}

\begin{table}[!ht]
    \centering
    \scriptsize
    \caption{Average LER and WER on monolingual test sets.}
    \begin{tabular}{lcccc}
    \toprule
        \multirow{2}{*}{\textbf{Method}} & \multicolumn{2}{c}{\bfseries Broadcast } & \multicolumn{2}{c}{\bfseries Institutional} \\ 
        & {LER} & {WER} & {LER} & {WER}  \\ 
    \midrule
        ECAPA-TDNN 107L & 17.08 & - & - & - \\ 
        ECAPA-TDNN 3L & 1.47 & - & 1.99 & - \\ 
        Monolingual ASR & - & 10.96 & - & 13.23 \\ 
        TitaNet-LID 3L & 0.75 & 11.42 & 2.39 & 15.06 \\ 
        Seg-SLI + VAD & 0.21 & 11.02 & 0.21 & 13.26 \\ 
        Seg-SLI + SD & 0.37 & 11.69 & 0.25 & 13.35 \\ 
        Frm-SLI + VAD & 0.26 & 11.06 & 0.35 & 13.40 \\
    \bottomrule
    \end{tabular}
\end{table}

We use the monolingual test sets to evaluate identification accuracy of the different SLI systems on domain-specific monolingual speech. A very good performance on monolingual data is crucial for applications where language changes are present but monolingual speech is predominant. For what concerns the broadcast data, we observe a decrease in performance for systems trained on VoxLingua when applied to L2 English. This loss is recovered with fine-tuning on the broadcast data, with all in-house systems having LER below 1\%. We observe a similar pattern for the institutional domain, with English (mostly L2 in this case) being the most critical language for the pretrained models, but not for the fine-tuned ones. We further explore the impact on transcription accuracy of the multilingual system composed by SLI and ASR, observing that monolingual ASR performance is only slightly degraded, thus proving the multilingual system to be a valid candidate for our use case.

\subsection{Multilingual Synthetic Speech}

\begin{table}[!ht]
    \centering
    \scriptsize
    \caption{LER of SLI systems on synthetic multilingual test set.}
    \begin{tabular}{lcccc}
    \toprule
        \multirow{2}{*}{\textbf{Method}} & \multicolumn{2}{c}{\bfseries ts-mix-short} & \multicolumn{2}{c}{\bfseries ts-mix-long} \\ 
        ~ & 1s gap & no gap & 1s gap & no gap \\ 
        \midrule
        ECAPA-TDNN 3L & 36.91 & 43.41 & 5.88 & 6.57 \\ 
        TitaNet-LID 3L & 1.69 & 36.90 & 0.20 & 5.72 \\ 
        Seg-SLI + VAD & 1.56 & 37.15 & 0.21 & 5.33 \\ 
        Seg-SLI + SD & 1.91 & 13.59 & 0.47 & 1.12 \\ 
        Frm-SLI + VAD & 0.73 & 4.57 & 0.08 & 0.51 \\ 
    \bottomrule
    \end{tabular}
\end{table}

On the multilingual synthetic datasets, the ECAPA-TDNN model has the highest LER on both short and long segments versions. This is mostly a consequence of the behaviour of the SpeechBrain VAD, that tends to create long segments containing multiple languages, that necessarily result in larger language confusion given the segment-based nature of the ECAPA-TDNN. We observe a similar performance for the TitaNet-LID and our segment-based model, when applied on VAD-extracted segments. In this case, the major impact on performance happens when there is no pause at the language change ("no gap" columns), because the VAD lacks cues to split the audio. This shortcoming was greatly reduced when using SD-based segmentation, because in this artificial scenario, similarly to our real use cases, language changes are associated to speaker changes. The best performing system on this synthetic testset is the frame-base SLI, that, given the same VAD, reduces the segment-based system error from 37.15\% to 4.57\% on the short test set and from 5.33\% to 0.51\% on the long test set. However, we notice that the test conditions (i.e. short segments of different languages stitched one after the other) are very close to the training conditions for the frame-based SLI, thus this system might be implicitly advantaged. 

\subsection{Real Multilingual Speech}

\begin{table}[!ht]
    \centering
    \scriptsize
    \caption{LDER and WER of SLI systems on multilingual test sets. 95\% confidence intervals are within squared brackets. Only WERs of fully integrated SLI+ASR systems are reported.}
    \label{tab:res_multi_real}
    \begin{tabular}{lccccc}
    \multicolumn{6}{l}{\em(a) Broadcast test set.}\\
    \toprule
        \textbf{Method} & \textbf{LDER} & \textbf{LC} & \textbf{MS} & \textbf{FA} & \textbf{WER} \\ 
    \midrule
        ECAPA-TDNN 3L & 12.13 [12.11-12.15] & 5.91 & 1.26 & 4.96 & - \\ 
        TitaNet-LID 3L & 14.09 [14.06-14.11] & 4.98 & 1.10 & 8.01 & 23.20 \\ 
        Seg-SLI + VAD & 13.83 [13.80-13.85] & 4.72 & 1.10 & 8.01 & 23.09 \\ 
        Seg-SLI + SD & 10.87 [10.84-10.89] & 1.57 & 1.43 & 7.86 & 21.19 \\ 
        Frm-SLI + VAD & 13.15 [13.13-13.17] & 4.13 & 1.00 & 8.02 & 23.12 \\ 
    \bottomrule
    \end{tabular}
    
    \smallskip
    \begin{tabular}{lccccc}
    \multicolumn{6}{l}{\em(b) Institutional test set.}\\
    \toprule
        \textbf{Method} & \textbf{LDER} & \textbf{LC} & \textbf{MS} & \textbf{FA} & \textbf{WER} \\ 
    \midrule
        ECAPA-TDNN 3L & 13.40 [13.32-13.47] & 7.46 & 2.52 & 3.42 & - \\ 
        TitaNet-LID 3L & 13.57 [13.50-13.57] & 8.32 & 2.86 & 2.39 & 26.29 \\ 
        Seg-SLI + VAD & 7.00 [6.95-7.05] & 1.75 & 2.86 & 2.39 & 21.18 \\ 
        Seg-SLI + SD & 6.47 [6.42-6.53] & 1.61 & 2.67 & 2.19 & 21.50 \\ 
        Frm-SLI + VAD & 6.84 [6.79-6.89] & 1.70 & 2.73 & 2.40 & 21.17 \\ 
    \bottomrule
    \end{tabular}
\end{table}

Table \ref{tab:res_multi_real} shows the performance of the SLI systems on the authentic multilingual test sets. 

Concerning the broadcast test set, the system combining SD and SLI reaches the lowest LDER. This is mostly due to a LC value (1.57\%) much lower than the other systems. The ECAPA-TDNN shows the highest LC, most likely due to the SpeechBrain VAD doing fewer splits than our in-house VAD. On the other hand, the SpeechBrain VAD performs best in terms of false alarms. 
Despite being the best system on synthetic multilingual data, the frame-based SLI does not significantly improve the performance of a segment-based SLI, neither in terms of LDER nor WER, significantly underperforming the system combining SD and SLI on the broadcast testset. By qualitative evaluations of the SLI systems predictions, we observe that the language boundaries are detected more sharply by the system combining SD and SLI than from the frame-based SLI. We hypothesize that this is a consequence of language-distinctive features happening at a longer timescale with respect to speaker-distinctive features, that result in a more accurate detection of speaker boundaries with respect to language boundaries. In addition, we observe that the frame-based system tends to over-segment when output language probabilities are very close to one another, having an impact on both LDER and WER. Lastly, the presence of English words in the Swedish speech might also be a source of errors for the frame-based SLI.

Regarding the institutional test set, we observe a big difference between LDER of the pretrained systems and our in-house systems. This is mostly due to much higher language confusion, that directly relates to the higher LER on the monolingual testset.
The best performing system in terms of LDER is again the system combining SD and SLI, even though, in this particular scenario, by a small margin w.r.t. the system employing a VAD. This is mostly due to the specific characteristics of the test set. Here, language (speaker) turns are well separated, often by relatively long pauses, and overlaps are very uncommon. In this case, the VAD successfully detects language boundaries, so there is not much room for improvement for other systems.
We report WER on this dataset for thoroughness, but values should be considered only as indicative, because the official transcripts of this dataset are not verbatim. 

\section{Conclusions}

We analyzed the performance of different spoken language identification systems on multilingual broadcast and institutional speech. These domains are characterized by language changes that happen at relatively long timescales and usually match speaker changes. We found that, under these circumstances, a cascaded system composed of speaker diarization and segment-based SLI is particularly accurate in detecting language changes, outperforming systems that are agnostic to speaker information.  
Furthermore, we analyzed the impact on transcription accuracy when the spoken language identifier is combined with monolingual speech recognition engines. We found that such a system has the lowest word error rate on multilingual data, with negligible degradation of monolingual transcription accuracy.

\bibliographystyle{IEEEtran}
\bibliography{mybib}

\begin{thebibliography}{10}
\providecommand{\url}[1]{#1}
\csname url@samestyle\endcsname
\providecommand{\newblock}{\relax}
\providecommand{\bibinfo}[2]{#2}
\providecommand{\BIBentrySTDinterwordspacing}{\spaceskip=0pt\relax}
\providecommand{\BIBentryALTinterwordstretchfactor}{4}
\providecommand{\BIBentryALTinterwordspacing}{\spaceskip=\fontdimen2\font plus
\BIBentryALTinterwordstretchfactor\fontdimen3\font minus \fontdimen4\font\relax}
\providecommand{\BIBforeignlanguage}[2]{{%
\expandafter\ifx\csname l@#1\endcsname\relax
\typeout{** WARNING: IEEEtran.bst: No hyphenation pattern has been}%
\typeout{** loaded for the language `#1'. Using the pattern for}%
\typeout{** the default language instead.}%
\else
\language=\csname l@#1\endcsname
\fi
#2}}
\providecommand{\BIBdecl}{\relax}
\BIBdecl

\bibitem{li2013}
H.~Li, B.~Ma, and K.~A. Lee, ``{Spoken Language Recognition: From Fundamentals to Practice},'' \emph{Proceedings of the IEEE}, vol. 101, no.~5, pp. 1136--1159, 2013.

\bibitem{valk2020voxlingua}
J.~Valk and T.~Alumäe, ``{VOXLINGUA107: A Dataset for Spoken Language Recognition},'' in \emph{Proc. of Spoken Language Technology Workshop}.\hskip 1em plus 0.5em minus 0.4em\relax IEEE, 2021, pp. 652--658.

\bibitem{ardila2020}
R.~Ardila, M.~Branson, K.~Davis, M.~Henretty, M.~Kohler, J.~Meyer, R.~Morais, L.~Saunders, F.~M. Tyers, and G.~Weber, ``{Common Voice: A Massively-Multilingual Speech Corpus},'' in \emph{Proc. of Language Resources and Evaluation Conference}.\hskip 1em plus 0.5em minus 0.4em\relax European Language Resources Association, 2020, pp. 4211--4215.

\bibitem{conneau2023}
A.~Conneau, M.~Ma, S.~Khanuja, Y.~Zhang, V.~Axelrod, S.~Dalmia, J.~Riesa, C.~Rivera, and A.~Bapna, ``{FLEURS: FEW-Shot Learning Evaluation of Universal Representations of Speech},'' in \emph{Proc. of Spoken Language Technology Workshop}.\hskip 1em plus 0.5em minus 0.4em\relax IEEE, 2023, pp. 798--805.

\bibitem{lyu2013}
D.-C. Lyu, E.-S. Chng, and H.~Li, ``{Language Diarization for Code-Switch Conversational Speech},'' in \emph{Proc. of International Conference on Acoustics, Speech and Signal Processing}.\hskip 1em plus 0.5em minus 0.4em\relax IEEE, 2013, pp. 7314--7318.

\bibitem{barras2020}
C.~Barras, V.-B. Le, and J.-L. Gauvain, ``{Vocapia-LIMSI System for 2020 Shared Task on Code-switched Spoken Language Identification},'' in \emph{{Proc. of Speech Technologies for Code-Switching in Multilingual Communities Workshop}}, 2020.

\bibitem{liu21_interspeech}
H.~Liu, L.~P.~G. Perera, X.~Zhang, J.~Dauwels, A.~W. Khong, S.~Khudanpur, and S.~J. Styles, ``{End-to-End Language Diarization for Bilingual Code-Switching Speech},'' in \emph{Proc. of Interspeech}.\hskip 1em plus 0.5em minus 0.4em\relax ISCA, 2021, pp. 1489--1493.

\bibitem{jia23b_interspeech}
F.~Jia, N.~R. Koluguri, J.~Balam, and B.~Ginsburg, ``{A Compact End-to-End Model with Local and Global Context for Spoken Language Identification},'' in \emph{Proc. of Interspeech}.\hskip 1em plus 0.5em minus 0.4em\relax ISCA, 2023, pp. 5321--5325.

\bibitem{mishra23_interspeech}
J.~Mishra, J.~N. Patil, A.~Chowdhury, and M.~Prasanna, ``{End to End Spoken Language Diarization with Wav2vec Embeddings},'' in \emph{Proc. of Interspeech}.\hskip 1em plus 0.5em minus 0.4em\relax ISCA, 2023, pp. 501--505.

\bibitem{graves2005framewise}
A.~Graves and J.~Schmidhuber, ``Framewise phoneme classification with bidirectional lstm and other neural network architectures,'' \emph{Neural networks}, vol.~18, no. 5-6, pp. 602--610, 2005.

\bibitem{speechbrain}
M.~Ravanelli, T.~Parcollet, P.~Plantinga, A.~Rouhe, S.~Cornell, L.~Lugosch, C.~Subakan, N.~Dawalatabad, A.~Heba, J.~Zhong, J.-C. Chou, S.-L. Yeh, S.-W. Fu, C.-F. Liao, E.~Rastorgueva, F.~Grondin, W.~Aris, H.~Na, Y.~Gao, R.~D. Mori, and Y.~Bengio, ``{SpeechBrain}: A general-purpose speech toolkit,'' \emph{arXiv:2106.04624}, 2021.

\bibitem{han20_interspeech}
W.~Han, Z.~Zhang, Y.~Zhang, J.~Yu, C.-C. Chiu, J.~Qin, A.~Gulati, R.~Pang, and Y.~Wu, ``{ContextNet: Improving Convolutional Neural Networks for Automatic Speech Recognition with Global Context},'' in \emph{Proc. of Interspeech}.\hskip 1em plus 0.5em minus 0.4em\relax ISCA, 2020, pp. 3610--3614.

\bibitem{kinoshita21_interspeech}
K.~Kinoshita, M.~Delcroix, and N.~Tawara, ``{Advances in Integration of End-to-End Neural and Clustering-Based Diarization for Real Conversational Speech},'' in \emph{Proc. of Interspeech}.\hskip 1em plus 0.5em minus 0.4em\relax ISCA, 2021, pp. 3565--3569.

\bibitem{park2022review}
T.~J. Park, N.~Kanda, D.~Dimitriadis, K.~J. Han, S.~Watanabe, and S.~Narayanan, ``{A review of speaker diarization: Recent advances with deep learning},'' \emph{Computer Speech \& Language}, vol.~72, p. 101317, 2022.

\bibitem{fujita2019end}
Y.~Fujita, N.~Kanda, S.~Horiguchi, Y.~Xue, K.~Nagamatsu, and S.~Watanabe, ``End-to-end neural speaker diarization with self-attention,'' in \emph{Proc. of Automatic Speech Recognition and Understanding Workshop}.\hskip 1em plus 0.5em minus 0.4em\relax IEEE, 2019, pp. 296--303.

\bibitem{zeinali2019but}
H.~Zeinali, S.~Wang, A.~Silnova, P.~Mat{\v{e}}jka, and O.~Plchot, ``{BUT System Description to VoxCeleb Speaker Recognition Challenge 2019},'' \emph{arXiv:1910.12592}, 2019.

\bibitem{chung2020spot}
J.~S. Chung, J.~Huh, A.~Nagrani, T.~Afouras, and A.~Zisserman, ``Spot the conversation: speaker diarisation in the wild,'' in \emph{Proc. of Interspeech}.\hskip 1em plus 0.5em minus 0.4em\relax ISCA, 2020.

\bibitem{cieri2004fisher}
C.~Cieri, D.~Miller, and K.~Walker, ``{The Fisher corpus: A resource for the next generations of speech-to-text.}'' in \emph{Proc. of Language Resources and Evaluation Conference}, vol.~4.\hskip 1em plus 0.5em minus 0.4em\relax European Language Resources Association, 2004, pp. 69--71.

\bibitem{fujita2019bend}
Y.~Fujita, N.~Kanda, S.~Horiguchi, K.~Nagamatsu, and S.~Watanabe, ``End-to-end neural speaker diarization with permutation-free objectives,'' in \emph{Proc. of Interspeech}, vol. 2019.\hskip 1em plus 0.5em minus 0.4em\relax ISCA, 2019, pp. 4300--4304.

\bibitem{snyder2015musan}
D.~Snyder, G.~Chen, and D.~Povey, ``Musan: A music, speech, and noise corpus,'' \emph{arXiv:1510.08484}, 2015.

\bibitem{ko2017study}
T.~Ko, V.~Peddinti, D.~Povey, M.~L. Seltzer, and S.~Khudanpur, ``A study on data augmentation of reverberant speech for robust speech recognition,'' in \emph{Proc. of International Conference on Acoustics, Speech and Signal Processing}.\hskip 1em plus 0.5em minus 0.4em\relax IEEE, 2017, pp. 5220--5224.

\bibitem{kuchaiev2019nemo}
O.~Kuchaiev, J.~Li, H.~Nguyen, O.~Hrinchuk, R.~Leary, B.~Ginsburg, S.~Kriman, S.~Beliaev, V.~Lavrukhin, J.~Cook, P.~Castonguay, M.~Popova, J.~Huang, and J.~M. Cohen, ``{NeMo: a toolkit for building AI applications using Neural Modules},'' \emph{arXiv:1909.09577}, 2019.

\bibitem{wang2023wespeaker}
H.~Wang, C.~Liang, S.~Wang, Z.~Chen, B.~Zhang, X.~Xiang, Y.~Deng, and Y.~Qian, ``Wespeaker: A research and production oriented speaker embedding learning toolkit,'' in \emph{Proc. of International Conference on Acoustics, Speech and Signal Processing}.\hskip 1em plus 0.5em minus 0.4em\relax IEEE, 2023, pp. 1--5.

\bibitem{chung2018voxceleb2}
J.~S. Chung, A.~Nagrani, and A.~Zisserman, ``{VoxCeleb2: Deep speaker recognition},'' in \emph{Proc. of Interspeech}.\hskip 1em plus 0.5em minus 0.4em\relax ISCA, 2018.

\bibitem{chua23_interspeech}
V.~Y.~H. Chua, H.~Liu, L.~P. Garcia, F.~T. Woon, J.~Wong, X.~Zhang, S.~Khudanpur, A.~W.~H. Khong, J.~Dauwels, and S.~J. Styles, ``{MERLIon CCS Challenge: A English-Mandarin code-switching child-directed speech corpus for language identification and diarization},'' in \emph{Proc. of Interspeech}.\hskip 1em plus 0.5em minus 0.4em\relax ISCA, 2023, pp. 4109--4113.

\end{thebibliography}

\end{document}